\documentstyle[spie]{article} 
\input{psfig}   

\title{The Chandra X-Ray Observatory's Radiation Environment and the
AP-8/AE-8 Model} 

\author{Shanil N. Virani\supit{a}, Reinhold
M\"{u}ller-Mellin\supit{b}, Paul P. Plucinsky\supit{a}, and Yousaf
M. Butt\supit{a}  
\skiplinehalf 
\supit{a}Harvard-Smithsonian Center for Astrophysics,
Cambridge, MA  02138
\\
\supit{b}Extraterrestriche Physik, IEAP, Universit\"{a}t Kiel, 24118, Kiel, Germany
}

\authorinfo{Further author information: (Send correspondence to SNV) \\
SNV: E-mail: svirani@cfa.harvard.edu \\
PPP: E-mail: pplucinsky@cfa.harvard.edu \\
YMB: E-mail: ybutt@cfa.harvard.edu \\
RMM: E-mail: mueller-mellin@kernphysik.uni-kiel.de}
\begin{document} 
  \maketitle 

%%%%%%%%%%%%%%%%%%%%%%%%%%%%%%%%%%%%%%%%%%%%%%%%%%%%%%%%%%%%% 
\begin{abstract}
	The \textit{Chandra X-ray Observatory} (CXO) was launched on July 23, 
1999 and reached its final orbit on August 7, 1999. The CXO is in a highly
elliptical orbit, approximately 140,000 km x 10,000 km, and has a
period of approximately 63.5 hours ($\approx$ 2.65 days). It transits the 
Earth's Van Allen 
belts once per orbit during which no science observations can be performed due 
to the high radiation environment. The \textit{Chandra X-ray
Observatory Center} 
(CXC) currently uses the National Space Science Data Center's ``near Earth'' 
AP-8/AE-8 radiation belt model to predict the start and end times of passage
through the radiation belts. However, our scheduling software uses
only a 
simple dipole model of the Earth's magnetic field. The resulting B, L 
magnetic coordinates, do not always give sufficiently accurate predictions of
the start and end times of transit of the Van Allen belts.  We show this 
by comparing to the data from Chandra's on-board radiation monitor, the
\textit{EPHIN} (Electron, Proton, Helium Instrument particle detector)
instrument.  
We present evidence that demonstrates this mis-timing of the outer
electron radiation belt as well as data that also demonstrate the
significant variablity
of one radiation belt transit to the next as  experienced by the CXO. We 
also present an explanation for why the dipole implementation of the 
AP-8/AE-8 model is not ideally suited for the CXO. Lastly, we provide a brief
discussion of our on-going efforts to identify a model that accounts
for radiation belt variability, geometry, and one that can be used for 
observation scheduling purposes.
\end{abstract}

%>>>> Please include a list of keywords after the abstract 


%%%%%%%%%%%%%%%%%%%%%%%%%%%%%%%%%%%%%%%%%%%%%%%%%%%%%%%%%%%%%
\section{INTRODUCTION}
\label{sect:intro}  % \label{} allows reference to this section

Just past midnight on July 23, 1999, the space shuttle
\textit{Columbia} lifted-off from Cape Canaveral, Florida. In its payload
bay lay the \textit{Chandra X-ray Observatory} (CXO), the primary
cargo of the \textit{STS-93} mission. Just under 8 hours after 
launch, Chandra was
deployed from the space shuttle. However, it would be nearly two
weeks later, after an Inertial Upper Stage booster and several
``burns'' by its own propulsion system, that Chandra would reach its 
final orbit. The CXO is now the third of NASA's ``great
observatories'' in space.

The CXO's operational orbit has an apogee of approximately 140,000 km
and a perigee of nearly 10,000 km, with a $28.5^\circ$ initial inclination.
The CXO's highly elliptical orbit, with an orbital period of
approximately 2.65 days, results in high efficiency for
observing. Moreover, the fraction of the sky occulted by the Earth is
small over most of the orbital period, as is the fraction of time when
the detector backgrounds are high as the flight system dips into the
Earth's radiation belts. Consequently, approximately 85\% of Chandra's
orbit is available for observing. In fact, uninterrupted observations
lasting as long as 2.3 days are possible\cite{sodell98}.

The CXO carries two focal plane science instruments: the High
Resolution Camera (HRC) and the Advanced CCD Imaging Spectrometer
(ACIS). The Observatory also possesses two objective transmission
gratings: a Low Energy Transmission Grating (LETG) that is to be 
primarily used with the HRC, and the High Energy Transmission Grating 
(HETG) that is to be primarily used with the ACIS. In addition to
these instruments, Chandra also carries a radiation monitor -- the
\textit{Electron, Proton, Helium Instrument} (EPHIN) particle detector.

In order to attain this high level of observing efficiency, a robust 
radiation environment model is required so that times of high
radiation are known \textit{a priori} when producing a weekly sequence
of CXO observations. The CXO orbit encounters higher radiation levels
as the spacecraft approaches perigee. Routine science observing will
cease whenever the scheduling system has indicated that the maximum
radiation levels are excessive. 

These times of high radiation are required not only so that
observations do not take place but also so as to not
cause significant radiation damage to the CXO's focal plane
instruments over the expected on-orbit design life.
To that end, the \textit{Chandra X-ray Center} (CXO) 
currently uses the National Space Science Data Center's (NSSDC) 
``near Earth'' AP-8/AE-8 radiation belt model to predict the start 
and end times of passage through the radiation belts.

In this paper, we provide a brief synopsis of the CXO and its
instruments in Section~\ref{chandra}. In Section~\ref{ephin}, we
present an overview of the Earth's magnetosphere and the EPHIN
detector. The AP-8/AE-8 model is
described in Section~\ref{ap8}. The following section,
Section~\ref{data-fits}, presents our evidence that our
implementation of the AP-8/AE-8 model does not always give
sufficiently accurate predictions of the start and end times of
transit of the Earth's Van Allen belts. In that same section, we 
present an explanation for why the dipole implementation of the 
AP-8/AE-8 model gives inaccurate start and end times for radiation belt
transit 75\% of the time (for low energy electrons). 
Lastly, in Section~\ref{future}, we provide a brief summary of our 
current operating procedure and the on-going work being done by the 
CXC and NASA's Marshall Space Flight Center's Radiation Environment
working group in identifying a new radiation model to be used for 
scheduling purposes.

%%%%%%%%%%%%%%%%%%%%%%%%%%%%%%%%%%%%%%%%%%%%%%%%%%%%%%%%%%%%%
\section{CHANDRA X-RAY OBSERVATORY'S FOCAL PLANE INSTRUMENTS} 
\label{chandra}

The observatory consists of a spacecraft system and a
telescope/science-instrument payload. The spacecraft system provides
mechanical controls, thermal control, electric power,
communication/command/data management, and pointing and aspect
determination. This section, however, only briefly describes the two
focal plane instruments on-board the CXO, the HRC and the ACIS, since
the main emphasis in this paper is the Observatory's radiation
environment and not the instruments \textit{per se}. Nevertheless, the
\textit{AXAF Observatory Guide}\cite{obsguide} and the 
\textit{AXAF Science Instrument Notebook}\cite{SIN} contain a wealth
of information about the \textit{CXO} and its instruments. More in depth
discussions of the Chandra mission, spacecraft, other instruments and
subsystems are presented elsewhere. \cite{weisskopf95}$^,$ 
\cite{zombeck96}$^,$ \cite{markert94}$^,$ \cite{brinkman87}

%%-----------------------------------------------------------
\subsection{High Resolution Camera (HRC)} 
\label{HRC}

The High Resolution Camera, \textit{HRC}, is a microchannel plate
(MCP) instrument. It is comprised of two detector elements, a $\sim$ 
100 mm square optimized for imaging (HRC-I) and a $\sim$ 20 x 300 mm 
rectangular device optimized for the Low Energy Transmission Grating 
(LETG) Spectrometer readout (HRC-S).

The HRC has the highest spatial resolution imaging on Chandra --
$\leq$ 0.5
arcsec (FWHM) -- matching the High Resolution Mirror Assembly (HRMA)
point spread function most closely. The HRC energy range extends to
low energies, where the HRMA effective area is the greatest. HRC-I has
a large field of view (31 arcmin on a side) and is useful for imaging
extended objects such as galaxies, supernova remnants, and clusters of
galaxies as well as resolving sources in a crowded field. The HRC has
good time resolution (16 $\mu$sec), valuable for the analysis of bursts,
pulsars, and other time-variable phenomena and limited energy
discrimination, \textit{E/$\Delta$E} $\sim$ 1 ($<$ 1 keV). The HRC-S is 
used primarily for readout of the low-energy grating, LETG, for which 
its large format with many pixels gives high spectral resolution ($>$ 
1000, 40-60 $\AA$) and wide spectral coverage (3 - 160 $\AA$).

\subsection{Advanced CCD Imaging Spectrometer (ACIS)} 
\label{ACIS}

\textit{ACIS} is the Advanced CCD Imaging Spectrometer. It is
comprised of two arrays of CCDs, one optimized for imaging wide
fields (2x2 chip array; ACIS-I), the other optimized for grating 
spectroscopy and for imaging narrow fields (1x6 chip array; ACIS-S).
Each array is shaped to follow the relevant focal surface. In 
conjunction with the HRMA, the ACIS imaging array provides
simultaneous
time-resolved imaging and spectroscopy in the energy range $\sim$ 
0.5 - 10.0 keV. When used in conjunction with the High Energy
Transmission Gratings (\textit{HETG}), the ACIS spectroscopic array
will acquire high resolution (up to \textit{E/$\Delta$E} = 1000)
spectra of point sources\cite{proposers}. The CCDs have an intrinsic 
energy resolution (\textit{E/$\Delta$E}) which varies from $\sim$5 to 
$\sim$50 across the energy range.

ACIS employs two varieties of CCD chips. Most of the chips are
``front-side'' (or FI) illuminated. That is, the front-side gate
structures are
facing the incident X-ray beam from the HRMA. Two of the 10 chips (S1
and S3) have had treatments applied to the back-sides of the chips,
removing the insensitive, undepleted, bulk silicon material and
leaving only the photo-sensitive depletion region exposed. These
``back-side'', or BI chips, are deployed with the back side facing the
HRMA. BI chips have a substantial improvement in low-energy quantum
efficiency as compared to the FI chips because no X-rays are lost to
the insensitive gate structures but suffer from poorer charge transfer
inefficiency, poorer spectral resolution, and poorer calibration
accuracy. In addition, early analysis from on-orbit data indicate that
the BIs are more susceptible to ``background flares'', which may
compromise a measurement, than are FIs\cite{plucinsk2000}. These
background flares (i.e. rapid increases in detector background) are
thought to be a consequence of \textit{Chandra's}
radiation environment.

\section{THE EARTH'S MAGNETOSPHERE AND THE EPHIN DETECTOR} 
\label{ephin}

Before describing the EPHIN detector, we will first provide a brief
overview of the Earth's magnetosphere.

\subsection{The Earth's Magnetosphere}

\begin{figure}
%>>>> following adds vertical space needed for figure; 
%  uncomment if figure is to be pasted into manuscript
%%   \vspace{7.5cm}
   \begin{center}
   \begin{tabular}{c}
   \psfig{figure=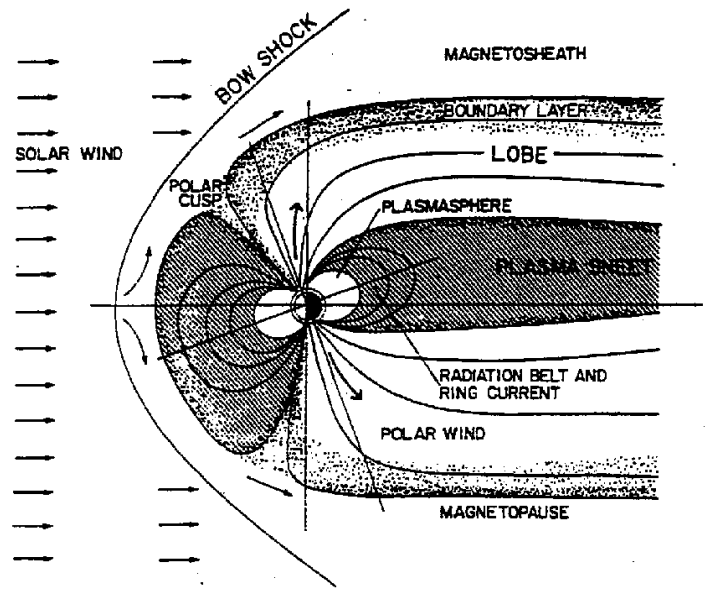,height=8cm} 
   \end{tabular}
   \end{center}
   \caption[example] 
%>>>> use \label inside caption to get Fig. number with \ref{}
   { \label{magneto}	Structure of the magnetosphere (Crooker and
   Siscoe, 1986)\cite{crook}  } 
   \end{figure}

The solar wind, a streaming ionized plasma, flows at all times and in
all directions from the Sun. When it encounters a magnetized obstacle,
the Earth for example, a magnetosphere is formed (see Figure 
~\ref{magneto}). The magnetosphere of a planet is defined as the
region where the particle motion is determined by the magnetic field
of the planet\cite{mueller}. The magnetopause is the boundary layer
which separates this region from the solar wind plasma. At the bow
shock, the solar wind, when sensing the obstacle prior to reaching the
magnetopause, undergoes an abrupt transition from supersonic flow to
subsonic flow. This shock allows the wind to be slowed, heated, and
deflected around the planet in the magnetosheath. The polar cusps of
clefts are singular points at northern and southern polar latitudes
where the magnetic field is zero. Only here can solar wind particles
directly reach the top of the atmosphere. Field lines in the
neighbourhood are either closed toward the dayside or open toward the
nightside, where the magnetotail is formed. The dayside extension of
the magnetosphere varies between 4.5 and 20 $R_{Earth}$ depending on the
solar wind pressure, whereas the nightside extension reaches several
hundred $R_{Earth}$, much farther than the orbit of the Moon. The
plasmasphere is a region of high density ($\sim$ $10^3$ $cm^{-3}$), cold
($\sim$ 1 eV) plasma, an extension of the ionosphere to altitudes up
to 3-4 $R_{Earth}$.

\begin{figure}
%>>>> following adds vertical space needed for figure; 
%  uncomment if figure is to be pasted into manuscript
%%   \vspace{7.5cm}
   \begin{center}
   \begin{tabular}{c}
   \psfig{figure=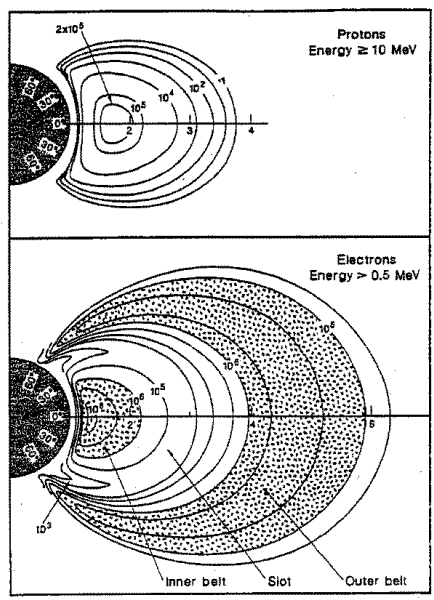,height=8cm} 
   \end{tabular}
   \end{center}
   \caption[example] 
%>>>> use \label inside caption to get Fig. number with \ref{}
   { \label{belts} Earth's Radiation Belts. The top panel shows the
   contours of the omnidirectional flux (counts/($cm^2$ s)) of protons
   with energies greater than 10 MeV. The bottom panel shows the
   contours of the omnidirectional flux of electrons with energies
   greater than 0.5 MeV. (Kivelson and Russell, 1995)\cite{kivel}} 
   \end{figure}

The region in which science observations must \textit{not} occur is in
the Van
Allen belts. The radiation belts, or the Van Allen belts, consist of
particles in orbits that circle the Earth from about 1,000 km above
the surface to a geocentric distance of $\sim$ 6 $R_{Earth}$. Contrary
to particles in the plasmasphere, these particles are trapped at high
energies. These particles enter the radiation belts through a variety
of means, including radial diffusion from more distant regions with
accompanying acceleration, and the decay of neutrons from the
sputtering of the atmosphere by galactic cosmic rays. Whereas
energetic protons form a single belt, as illustrated in the top panel
of Figure ~\ref{belts}, electrons form two belts -- an inner and an
outer belt (as illustrated in the bottom panel). A co-ordinate system
with \textit{L} (\textit{L} designates the magnetic-drift shell and is
equal to the distance in $R_{Earth}$ from the center of the Earth to
the point where the field line crosses the equator) and \textit{B}, 
the magnetic field, organizes the radiation belt data quite 
well\cite{mueller}. 

%-------------
%-------------

%-------------
%-------------

\subsection{The EPHIN Detector}

The natural space environment can cause a range of problems for a
spacecraft and may even compromise its mission. Environmental factors
include the radiation belts, solar energetic particles, cosmic rays,
plasma, gases, and micrometeorites. In this paper, we will only
address the radiation belts.

The CXO radiation environment is monitored by the EPHIN 
detector (although the HRC anti-coincidence shield may also be used
to some degree). The EPHIN was designed and manufactured as a
scientific instrument for the Solar and Heliospheric Observatory and
the Chandra X-ray Observatory by the University of Kiel. A detailed
description of the EPHIN instrument can be found in
M\"{u}ller-Mellin \textit{et al.} (1995)\cite{mueller2} and Sierks
(1997)\cite{sierks}.

\begin{figure}
%>>>> following adds vertical space needed for figure; 
%  uncomment if figure is to be pasted into manuscript
%%   \vspace{7.5cm}
   \begin{center}
   \begin{tabular}{c}
   \psfig{figure=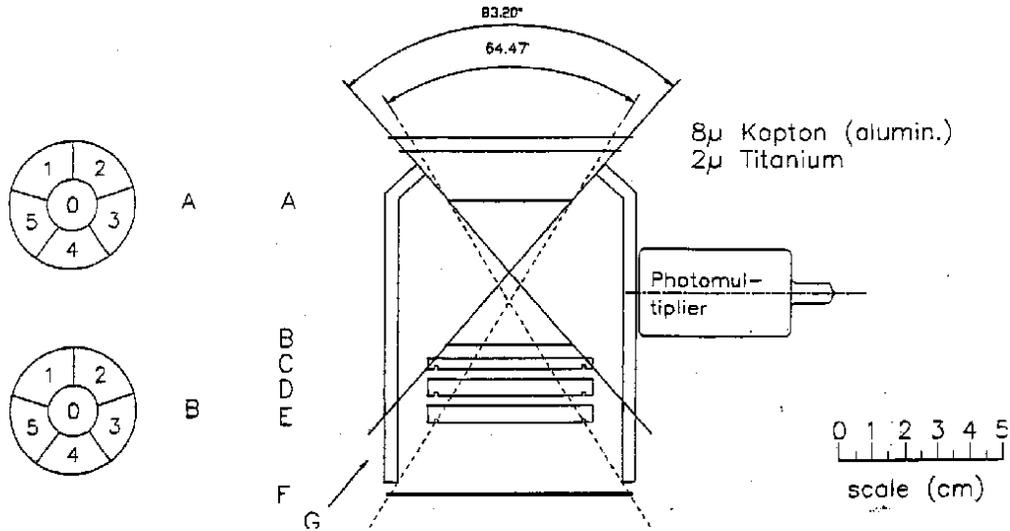,height=7cm} 
   \end{tabular}
   \end{center}
   \caption[example] 
%>>>> use \label inside caption to get Fig. number with \ref{}
   { \label{ephin-schem} EPHIN sensor schematics with detector A and B
   segmentation pattern on the left. The geometric factor varies with
   penetration depth (solid lines: view cone for particles stopping in
   B, dashed lines: view cone for penetrating particles). \cite{mueller3}} 
   \end{figure}

Briefly, the EPHIN consists of an array of 5 silicon detectors
with anti-coincidence to measure the energy of electrons in the range 150
keV - 10 MeV, and hydrogen/helium isotopes in the energy range 5 - 53
MeV/nucleon. Please see Table ~\ref{ephintable} for a more precise
overview of EPHIN's scientific data channels. As 
Figure ~\ref{ephin-schem} illustrates, the field of view
of EPHIN is $\sim$ 83 degrees. The stack of 5 silicon detectors
operates in a multi-\textit{dE/dx}
vs. \textit{E} mode and it is surrounded by an anti-coincidence shield
(Figure ~\ref{ephin-schem}). Two passivated ion-implanted silicon
detectors (PIPS) A and B define the 83 degree field of view with a
geometric factor of 5.1 $cm^2$ sr. The silicon detectors are divided
into 6 segments. This coarse position sensing permits sufficient
correction for path length variations needed to resolve isotopes of
hydrogen and helium. The directional information cannot be used for
anisotropy measurements as particles from different directions are
summed when they have the same degree of oblique incidence. Another
important advantage of this sector paradigm is the capability to
implement a commandable or self-adaptive geometric factor
\cite{mueller3}. This permits measurements of fluxes as high as 2 x
$10^5$ counts/($cm^2$ s sr) without significant dead time losses. A
third benefit is the reduction in capacitance at the input to the
charge sensitive pre-amplifiers, resulting in low-noise performance.

Lastly, the lithium-drifted silicon detectors (Si(Li)) C, D, and E
stop electrons up to 10 MeV and hydrogen and helium nuclei up to 53
MeV/n. The ion-implanted detector F will allow particles stopping in
the telescope to be distinguished from penetrating particles. The fast
plastic scintillation detector G, viewed by a 1-inch photomultiplier
and used in anti-coincidence, is indispensible for accurate electron
measurements. This whole stack of detectors is mounted in an aluminum
housing, the aperature being covered by a 2 $\mu$m thin titanium foil for
light tightness. A second foil in the viewing cone is made of 76 $\mu$m
aluminized Kapton (original specifications in Figure
~\ref{ephin-schem} called for 8 $\mu$m) for thermal control\cite{mueller3}.

\begin{table} [ht]   %>>>> [h] means place table here
\caption{EPHIN Energy Ranges\cite{mueller}} 
\label{ephintable}
\begin{center}       
\begin{tabular}{lcc} %% this creates two columns
%% |l|l| to left justify each column entry
%% |c|c| to center each column entry
%% use of \rule[]{}{} below opens up each row
\hline
\rule[-1ex]{0pt}{3.5ex}  EPHIN Channel & Energy Range & Energy Width \\
\rule[-1ex]{0pt}{3.5ex}                & [MeV] or [MeV/n] & [MeV] or [MeV/n] \\
\hline\hline
\rule[-1ex]{0pt}{3.5ex}  E150  & 0.25 - 0.70  & 0.45   \\
\rule[-1ex]{0pt}{3.5ex}  E300  & 0.67 - 3.00  & 2.3   \\
\rule[-1ex]{0pt}{3.5ex}  E1300 & 2.64 - 6.18  & 3.6 \\
\rule[-1ex]{0pt}{3.5ex}  E3000 & 4.80 - 10.4  & 5.6 \\
\hline
\rule[-1ex]{0pt}{3.5ex}  P4    & 5.0 - 8.3    & 3.3 \\
\rule[-1ex]{0pt}{3.5ex}  P8    & 8.3 - 25.0   & 16.7 \\
\rule[-1ex]{0pt}{3.5ex}  P25   & 25.0 - 41.0  & 16.0 \\
\rule[-1ex]{0pt}{3.5ex}  P41   & 41.0 - 53.0  & 12.0 \\
\hline
\rule[-1ex]{0pt}{3.5ex}  H4    & 5.0 - 8.3    & 3.3  \\
\rule[-1ex]{0pt}{3.5ex}  H8    & 8.3 - 25.0   & 16.7 \\
\rule[-1ex]{0pt}{3.5ex}  H25   & 25.0 - 41.0  & 16.0 \\
\rule[-1ex]{0pt}{3.5ex}  H41   & 41.0 - 53.0  & 12.0 \\
\hline
\rule[-1ex]{0pt}{3.5ex}  INT   & e: $>$ 8.7   & \\
\rule[-1ex]{0pt}{3.5ex}        & p,h: $>$ 53  & \\
\hline\hline
\end{tabular}
\end{center}
\end{table}

%-------------
%-------------

%%-------------
%   \begin{figure}
%%>>>> following adds vertical space needed for figure; 
%%  uncomment if figure is to be pasted into manuscript
%%%   \vspace{7.5cm}
%   \begin{center}
%   \begin{tabular}{c}
%   \psfig{figure=cxo.ps,height=10cm} 
%   \end{tabular}
%   \end{center}
%   \caption[example] 
%%>>>> use \label inside caption to get Fig. number with \ref{}
%   { \label{cxo} Mapping of the CXO orbit, magnetopause, and bow shock
%   in the ecliptic plane as the Earth revolves around the Sun\cite{mueller}.} 
%   \end{figure} 
%%-------------

\section{AP/AE TRAPPED PARTICLE RADIATION MODEL} 
\label{ap8}

The CXO's orbital parameters were designed to minimise the
time spent in the Van Allen radiation belts. Energetic particles
contribute to the background rate and can damage the CCD detectors
over time. Hence, the apogee is $\sim$ 140,000 km. Now, the time spent
above 60,000 km is a good indicator of the time spent outside the
radiation belts, however in order to maximize orbital efficiency, a
robust radiation environment model is required; the AP-8/AE-8
radiation belt model is the one the CXC has implemented to accomplish
this task. The primary outputs from the AE-8/AP-8 model are the
spatial fluxes of trapped electrons and protons in the near Earth
environment.

These maps contain omnidirectional, integral electron (AE maps) and 
proton (AP maps) fluxes in the energy range 0.04 MeV to 7 MeV for
electrons and 0.1 MeV to 400 MeV for protons in the Earth's radiation 
belt (L = 1.2 to 11 for electrons, L = 1.17 to 7 for protons)\cite{vette}. 
The fluxes are stored as functions of energy, L-value, and B/$B_0$ 
(magnetic field strength normalized to its equatorial value
on the field line). These maps are based on data from more than 20 
satellites that operated from the early sixties to the mid-seventies. 
Indeed, the data contained within these flux maps span 14 years (1966
to 1980). The AE-8/AP-8 model is the latest edition in a series of 
updates starting with AE-1 and AP-1 in 1966\cite{vette}. 

The different previous electron models can be characterised as inner 
(L = 1.2-3) or outer (L = 3-11) zone models and as models for solar 
cycle maximum or minimum conditions. However, AE-8 is the 
first model that covers the whole L range and both solar cycle
extrema. The AP maps differ in energy range and solar cycle
phase. AP-8 is the first model for the whole energy range and both
solar cycle extrema\cite{vette}. 

However, one significant shortcoming of the AP-8/AE-8 model is that
none of the flux maps consider time variations beyond the solar cycle 
minimum/maximum distinction. A study carried out by the NSSDC\cite{teague} 
found that for the inner zone electrons, the two dominant time effects
are caused by magnetic storms and the solar cycle. In particular, 
magnetic storms strongly affect electrons with energies higher than
0.7 MeV at higher L-shells, whereas the solar cycle effect is most 
significant for electrons with energies below 0.7 MeV. It is important
to note that magnetic storm effects are still not yet included in any
of the AE maps. It is thought that the largest errors occur where 
steep gradients in spatial and spectral distribution exist and where 
time variations are not well understood\cite{vette}.  Therefore, it 
is clear that one has to be careful in extrapolating these models
to later epochs. 

The electron AE-8 and proton AP-8 flux maps for solar maximum and
minimum are available through the NSSDC as part of its ``RADBELT
1988'' software program (see {\tt http://nssdc.gsfc.nasa.gov/}).
The software provides omnidirectional, integral (or differential) 
electron or proton fluxes in the Earth's radiation belt by using an 
interpolation procedure that uses the AE-8 and AP-8 trapped particle
flux maps. It is the output from this software program that we compare
against EPHIN data in the next section.

\section{AP-8/AE-8 MODEL VS. EPHIN DATA} 
\label{data-fits}

In order to make a meaningful comparison of the AP-8/AE-8 model with
the EPHIN data, the correct information must be provided to the radiation
belt software program. Since EPHIN possesses 4 electron channels and 4
proton channels (see Table ~\ref{ephintable}), the NSSDC software
routine was queried to find the start times when various specified
thresholds would be exceeded, and the end times when flux predictions 
would be below the same thresholds for each EPHIN channel. These time
spans are then, effectively, a proxy for radiation belt transit 
(as measured by the EPHIN).

Now, the offline scheduling system (OFLS), which incorporates the 
radiation belt
software that the CXC uses to produce transit predictions, is not
designed to search for specific ranges of energies for flux
excursions. That is, it can only do single energies with multiple flux
values; a single time is generated at each entry and exit where the
flux and energy conditions are met. Below is a sample query that 
was used to determine start and end times of radiation belt transit. 

{\bf FOR EPHIN ELECTRONS:}

\begin{description}
\item{E150:}  Flux =   1 cts/s/$cm^2$/sr  Energy $\geq$ 0.25 MeV 
\item{E300:}  Flux =   1 cts/s/$cm^2$/sr  Energy $\geq$ 0.67 MeV 
\item{E1300:} Flux =   1 cts/s/$cm^2$/sr  Energy $\geq$ 2.64 MeV 
\item{E3000:} Flux =   1 cts/s/$cm^2$/sr  Energy $\geq$ 4.80 MeV 
\end{description}

{\bf FOR EPHIN PROTONS:}

\begin{description}
\item{P4:}    Flux = 4.93 cts/s/$cm^2$/sr  Energy $\geq$  5.0 MeV
\item{P8:}    Flux =    1 cts/s/$cm^2$/sr  Energy $\geq$  8.3 MeV
\item{P25:}   Flux =    1 cts/s/$cm^2$/sr  Energy $\geq$ 25.0 MeV
\item{P41:}   Flux =    1 cts/s/$cm^2$/sr  Energy $\geq$ 41.0 MeV
\end{description}

The time spans (i.e., entries and exits) for which the above
conditions were met have been determined since Day 200 of 1999 to
the middle of February, 2000. The time spans, which
corresponds to an EPHIN proton or electron channel, have been
superimposed on the EPHIN data. Generally, the agreement between the
radiation belt model and the EPHIN data is quite poor. For instance,
of the 73 radiation belt transits seen in the EPHIN E150 channel, 55
were poorly timed ($\sim$ 75\%). That is, of the 73 radiation belt 
transits investigated, nearly 75\% showed EPHIN exceedances of the 
flux criteria specified above \textit{before} the expected
exceedance from the AE-8/AP-8 model. Similar numbers for the E300,
E1300, and E3000 channels are 48 (66\%), 20 (27\%) and 20 (27\%), 
respectively.

\begin{figure}
%>>>> following adds vertical space needed for figure; 
%  uncomment if figure is to be pasted into manuscript
%%   \vspace{7.5cm}
   \begin{center}
   \begin{tabular}{c}
   \psfig{figure=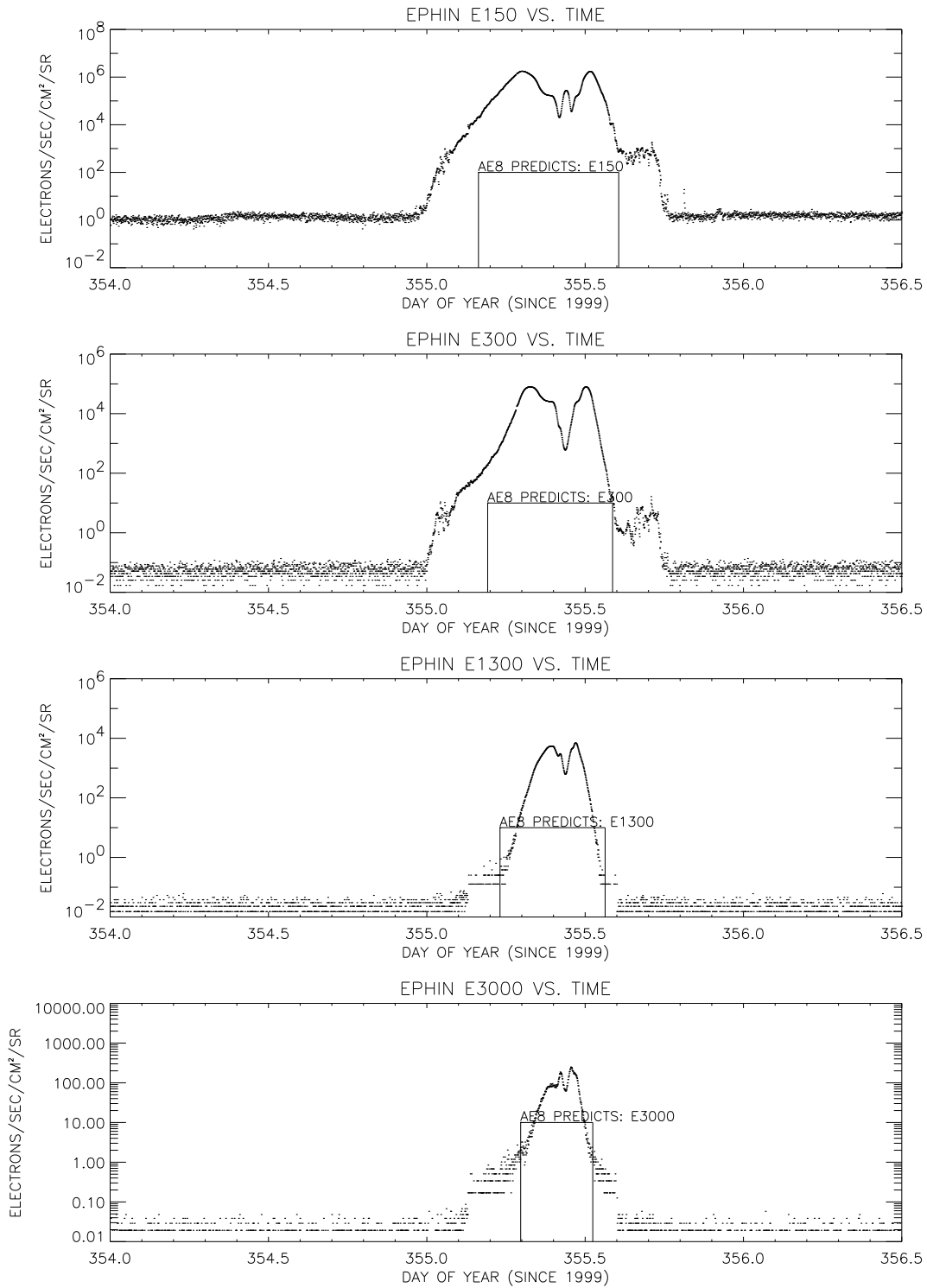,height=20cm} 
   \end{tabular}
   \end{center}
   \caption[example] 
%>>>> use \label inside caption to get Fig. number with \ref{}
   { \label{EPHIN-bad1}	  
AE-8 radiation belt model predictions against EPHIN data on Day 355. 
Note AE-8 predictions are based on an integral above an energy
threshold (specified within the text) whereas the EPHIN electron channel
has a much narrower energy window.} 
   \end{figure}

\begin{figure}
%>>>> following adds vertical space needed for figure; 
%  uncomment if figure is to be pasted into manuscript
%%   \vspace{7.5cm}
   \begin{center}
   \begin{tabular}{c}
   \psfig{figure=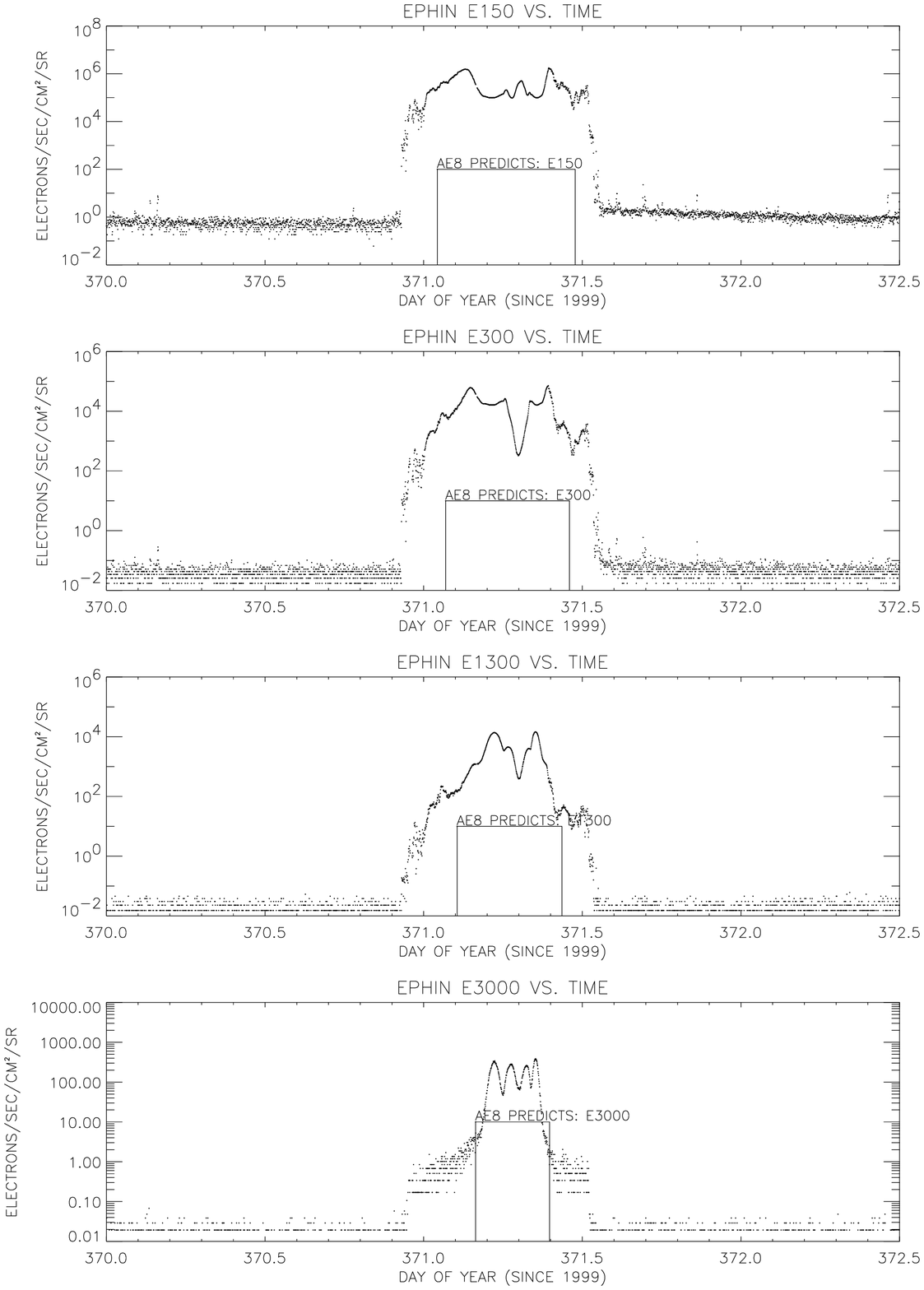,height=20cm} 
   \end{tabular}
   \end{center}
   \caption[example] 
%>>>> use \label inside caption to get Fig. number with \ref{}
   { \label{EPHIN-bad2}	  
AE-8 radiation belt model predictions against EPHIN data on Day 371. 
Note AE-8 predictions are based on an integral above an energy
threshold (specified within the text) whereas the EPHIN electron channel
has a much narrower energy window.} 
   \end{figure}

\begin{figure}
%>>>> following adds vertical space needed for figure; 
%  uncomment if figure is to be pasted into manuscript
%%   \vspace{7.5cm}
   \begin{center}
   \begin{tabular}{c}
   \psfig{figure=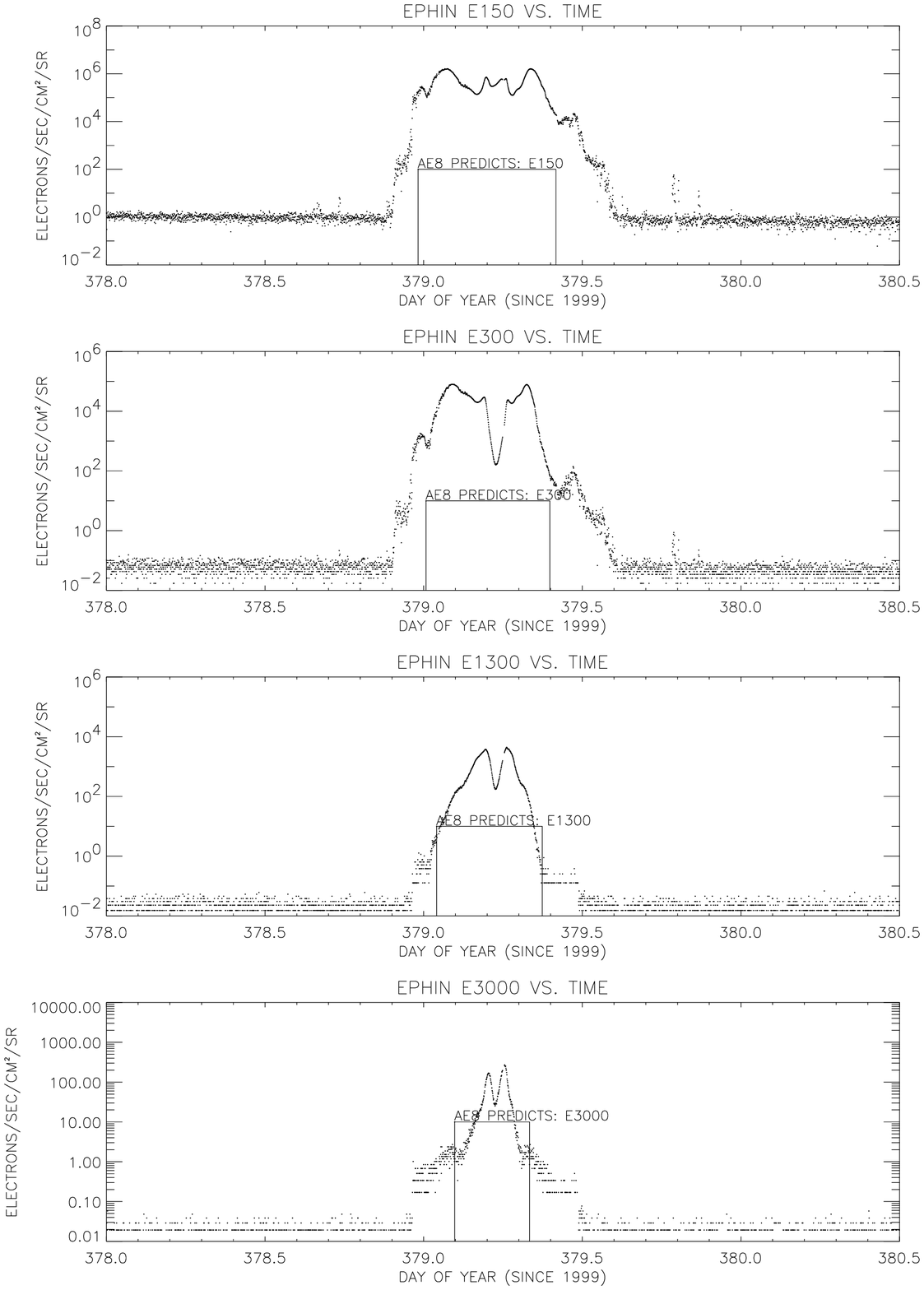,height=20cm} 
   \end{tabular}
   \end{center}
   \caption[example] 
%>>>> use \label inside caption to get Fig. number with \ref{}
   { \label{EPHIN-bad3}	  
AE-8 radiation belt model predictions against EPHIN data on Day 379.
Note AE-8 predictions are based on an integral above an energy
threshold (specified within the text) whereas the EPHIN electron channel
has a much narrower energy window. } 
   \end{figure}

Three of the most extreme examples of these radiation belt mis-timing 
are presented in Figures ~\ref{EPHIN-bad1}, ~\ref{EPHIN-bad2}, and 
~\ref{EPHIN-bad3}. Although similar data exists for the EPHIN proton 
channels, they are not included in this paper since it appears that 
in a high electron flux environment, electrons can mimic a proton
signal in the low energy EPHIN proton channels. More work is needed to
fully understand and solve this problem before that data is presented.

These plots also demonstrate the significant variability in 
radiation belt transits, however, one has to be very careful in
drawing conclusions about temporal variability during radiation belt
transits based on these plots alone since the EPHIN detector saturates
during perigee transit. The effect of the EPHIN anti-coincidence
counter, when in saturation, will distort the electron fluxes and time
profile in the innermost part of the radiation belts, approximately
when E150 rates exceed $10^5 counts/s/cm^2/sr$. The indicated
intensities in Figures ~\ref{EPHIN-bad1}, ~\ref{EPHIN-bad2},
and ~\ref{EPHIN-bad3} during
these intervals, should be regarded as lower limits only.
An internal CXC memo, however, used EPHIN's
leakage currents as a probe to understand Chandra's radiation
environment and also found significant variablity in radiation belt
intensity and duration\cite{shanil}. In the following table, we note
the average value of the discrepancy (between model and data) for each
EPHIN channel for the three perigee transits presented in Figures 
~\ref{EPHIN-bad1}, ~\ref{EPHIN-bad2}, and ~\ref{EPHIN-bad3}.

\begin{table} [ht]   %>>>> [h] means place table here
\caption{Average Time Differences Between EPHIN Electron Channels and the OFLS} 
\label{comp}
\begin{center}       
\begin{tabular}{lcc} %% this creates two columns
\hline
\rule[-1ex]{0pt}{3.5ex}  EPHIN Channel & $|T_{EPHIN}$-$T_{OFLS}$$|_{entry}$ & $|T_{EPHIN}$-$T_{OFLS}$$|_{exit}$ \\
\rule[-1ex]{0pt}{3.5ex}                &          ks         &       ks      \\
\hline\hline
\rule[-1ex]{0pt}{3.5ex}  E150  &  9.1  & 10.0 \\
\rule[-1ex]{0pt}{3.5ex}  E300  & 10.0  & 10.0 \\
\rule[-1ex]{0pt}{3.5ex}  E1300 &  4.3  &  2.4 \\
\rule[-1ex]{0pt}{3.5ex}  E3000 &  3.4  &  3.4 \\
\hline\hline
\end{tabular}
\end{center}
\end{table}

Hence, we observe that the predictive value of the AP-8/AE-8 model is 
adequate, though certainly not optimal. In particular, the mis-timing
of the softer species is seen to be worse than that for their harder 
counterparts. Since the degradation of the ACIS FI CCDs is believed to be
due to protons in the range $\sim$ 100 kev - $\sim$ 400 keV, it is important
for us to accurately predict the location of these relatively
low-energy species. Due to their low energy, however, these species are also
very dynamic, and can rapidly change their spatial extent subject to
the day-to-day solar conditions. In contrast, the implementation of 
AP-8/AE-8 model used is a static one and can only take into account solar 
MAX vs. MIN conditions, and we believe this is the major source of
the scheduling error of the radiation belts we witness, especially for
low-energy species. Other models, such as a 3-D version of the 
Magnetospheric Specification and Forecast Model\cite{msm}
currently under development will be better able to take into
account solar variations on a day-to-day (or week-to-week) basis and
will thus increase Chandra's observing efficiency by reducing the 
necessary ``padding'' of the scheduled radiation belts which is
currently implemented (discussed in Section ~\ref{future}) as a CXC
mission planning policy.

%%-------------
%   \begin{figure}
%%>>>> following adds vertical space needed for figure; 
%%  uncomment if figure is to be pasted into manuscript
%%%   \vspace{7.5cm}
%   \begin{center}
%   \begin{tabular}{c}
%   \psfig{figure=eD266.ps,height=23cm} 
%   \end{tabular}
%   \end{center}
%   \caption[example] 
%%>>>> use \label inside caption to get Fig. number with \ref{}
%   { \label{EPHIN-good}	  
%AE-8 radiation belt model predictions against EPHIN data on Day
%265.5. Note AE-8 predictions are based on an integral energy range
%whereas the EPHIN electron channel has a much narrower energy window.} 
%   \end{figure} 
%%-------------

%-------------
%-------------

%-------------
%-------------

%-------------
%-------------

\section{CXC OPERATING PROCEDURE AND FUTURE WORK} 
\label{future}

Although the ACIS hardware is fairly robust with respect to radiation,
repeated exposure to high levels of radiation will gradually degrade
the hardware (see ACIS Flight Software User's Guide). In fact,
shortly after opening the telescope door to celestial sources for the
first time, it was found that the FI ACIS CCDs had suffered radiation
damage ({\tt http://chandra.harvard.edu}). This radiation damage is
believed to have been caused by forward scattering of 
protons with energies between 100 keV and 400 keV, associated with
the radiation belts, from Chandra's mirrors
during passage through the Van Allen belts. Since this discovery, a
number of policy changes have been implemented that will hopefully
prevent further radiation damage from occuring.

Because of this mis-timing of the radiation belts by our scheduling
software, we have had to be more conservative in identifying times we
shut down the ACIS for the radiation belt transit and times we turn on
the ACIS post-perigee transit. From Table ~\ref{comp}, it can be seen 
that if a 10-11 ks ``pad'' was added to both sides of the predicted 
radiation belt times, one would get better agreement with the EPHIN
data. Indeed, a more thorough analysis of Chandra's radiation history
to date yielded the same result. Therefore, it was 
decided to pad the AE-8 predictions of radiation belt entry and exit 
by 13 ks. This is because the AE-8 radiation belt transit time span is
longer than the AP-8 time span since the proton belt lies below 
the outer electron belt (see Figure~\ref{belts}). Thus, this choice 
leads to the more conservative approach.

Additionally, since it is believed to be protons focused on
the ACIS as it sat in the focal plane during a series of radiation
belt transits early in the mission that has caused most of the radiation 
damage (and not this mis-timing of the ``wings'' of the radiation belt
itself), it was also decided to place the HRC in the focal plane 
(i.e., keep ACIS out from the focal plane) during subsequent radiation
belt transits and to have it partially close its door to protect its 
instrumentation. This translation of the science instruments occurs at
the ``modified'' start and end times of radiation belt transit.

To protect ACIS during the science operations portion of the orbit,
new threshold levels were uplinked to the CXO. Now, EPHIN data
is monitored by \textit{Chandra's} on-board computer (OBC) which will
activate commands to safe the focal plane instruments during periods
of high radiation (e.g. a solar flare or coronal mass ejection). All
of these new policy changes and our monitoring scheme will be
discussed in more depth in a forthcoming paper\cite{sodell2}.

However, clearly all of the above policies are ``short-term'' solutions. What
is really needed is a more robust, higher fidelity radiation belt model
that provides reliable start and end times of radiation belt
transit. To that end, work is currently in progress to produce an
empirical model of the Earth's radiation belt that will provide the
CXC with reliable perigee transit information\cite{blackwell}.

Lastly, because of the present radiation damage that has been
sustained by our FI ACIS CCDs, our radiation monitor, EPHIN,
has become of paramount importance in preserving the health and
safety of the CXO. Thus, work has been on-going to further analyse its
data, to better understand the saturation issues with perigee transit,
and to develop a procedure that is able to account for electron 
contamination of the proton data. Clearly, more work is needed so that
the outstanding science being returned by the 
\textit{Chandra X-ray Observatory} can continue unattenuated for many 
more years to come.

%%%%%%%%%%%%%%%%%%%%%%%%%%%%%%%%%%%%%%%%%%%%%%%%%%%%%%%%%%%%%
\acknowledgments     %>>>> equivalent to \section*{ACKNOWLEDGMENTS}       
 
We are grateful to many people for their support, encouragement,
fruitful discussions, suggestions, and data analysis. In particular, 
we would like to single out Stephen O'Dell and the entire CXC/Marshall 
Space Flight Center Radiation Environment team, Robert Cameron, 
Michael Juda, Richard Edgar, Dan Shropshire, and Dan Schwartz.  
The authors acknowledge support for this research from NASA contract 
NAS8-39073.

%%%%%%%%%%%%%%%%%%%%%%%%%%%%%%%%%%%%%%%%%%%%%%%%%%%%%%%%%%%%%
%%%%% References %%%%%

  \bibliography{report}   %>>>> bibliography data in report.bib

\begin{thebibliography}{10}

\bibitem{sodell98}
S.~L. O'Dell and M.~C. Weisskopf, ``Advanced {X}-ray {A}strophysics {F}acility
  ({AXAF}): {C}alibration {O}verview,'' in {\em X-{R}ay {O}ptics,
  {I}nstruments, and {M}issions},  R.~B. Hoover and A.~B.~W. II, eds., {\em
  Proc. SPIE} {\bf 3444}, pp.~2--18, 1998.

\bibitem{obsguide}
A.~T. 403, {\em AXAF Observatory Guide}, Chandra X-ray Science Center,
  Cambridge, MA, 1997.

\bibitem{SIN}
A.~T. 401, {\em {AXAF} {S}cience {I}nstrument {N}otebook}, Chandra X-ray
  Science Center, Cambridge, MA, 1995.

\bibitem{weisskopf95}
M.~C. Weisskopf, S.~L. O'Dell, and R.~F. Elsner, ``Advanced {X}-ray
  {A}strophysics {F}acility - {AXAF} an {O}verview,'' in {\em X-{R}ay and
  {E}xtreme {U}ltraviolet {O}ptics},  R.~B. Hoover and A.~B.~W. II, eds., {\em
  Proc. SPIE} {\bf 2515}, 1995.

\bibitem{zombeck96}
M.~V. Zombeck, ``Advanced {X}-ray {A}strophysics {F}acility {AXAF},'' in {\em
  Proceedings of the International School of Space Science Course on X-Ray
  Astronomy},  {\em Aquila, Italy} {\bf CfA Preprint 403}, 1996.

\bibitem{markert94}
T.~H. Markert, C.~R. Canizares, D.~Dewey, M.~McGuirk, C.~S. Pak, and M.~L.
  Schattenburg, ``{H}igh {E}nergy {T}ransmission {G}rating {S}pectrometer
  ({HETGS}) for {AXAF},'' in {\em {EUV}, {X-R}ay, and {G}amma-{R}ay
  {I}nstrumentation for {A}stronomy {V}},  {\em Proc. SPIE} {\bf 2280}, 1994.

\bibitem{brinkman87}
A.~C. Brinkman, J.~J. van Rooijen, J.~A.~M. Bleeker, J.~H. Dijkstra, J.~Heise,
  P.~A.~J. de~Korte, R.~Mewe, and F.~Paerels, ``{L}ow {E}nergy {X}-ray
  {T}ransmission {G}rating {S}pectrometer for {AXAF},'' {\em Astro. Lett.} {\bf
  26}, p.~73B, 1987.

\bibitem{proposers}
A.~T. 402, {\em {AXAF} {P}roposers' {G}uide}, Chandra X-ray Science Center,
  Cambridge, MA, 1997.

\bibitem{plucinsk2000}
P.~P. Plucinsky and S.~N. Virani, ``Observed {O}n-{O}rbit {B}ackground of the
  {ACIS} {D}etector on the {C}handra {X}-ray {O}bservatory,'' in {\em X-{R}ay
  {O}ptics, {I}nstruments, and {M}issions},  J.~Truemper and B.~Aschenbach,
  eds., {\em Proc. SPIE} {\bf 4012}, p.~(this volume), 2000.

\bibitem{mueller}
R.~Mueller-Mellin, {\em {EPHIN} {R}esponse to the {AXAF} {R}adiation
  {E}nvironment}, Doc. No. EPH-RRE-001, Kiel, Germany, 1997.

\bibitem{crook}
N.~U. Crooker and G.~L. Siscoe, ``{T}he {E}ffect of the {S}olar {W}ind on the
  {T}errestrial {E}nvironment,'' in {\em {P}hysics of the {S}un},  P.~Sturrock,
  T.~Holzer, D.~Mihalas, and R.~Ulrich, eds., vol.~III, 1986.

\bibitem{kivel}
M.~Kivelson and C.~Russell, {\em {I}ntroduction to {S}pace {P}hysics},
  Cambridge University Press, Cambridge, England, 1995.

\bibitem{mueller2}
R.~Mueller-Mellin, H.~Kunow, V.~Fleissner, E.~Pehlke, E.~Rode, N.~Roschmann,
  C.~Scharmberg, H.~Sierks, P.~Rusznyak, I.~E. S.~McKenna-Lawlor, J.~Sequeiros,
  D.~Meziat, S.~Sanchez, J.~Medina, L.~der Peral, M.~Witte, R.~Marsden, and
  J.Henrion, ``Costep - comprehensive suprathermal and energetic particle
  analyser,'' {\em Solar Physics} {\bf 162}, pp.~483--504, 1995.

\bibitem{sierks}
H.~Sierks, {\em Kosmische Teilchen im Sonnensystem - Messung Geladener Teilchen
  mit dem Kieler Instrument EPHIN an Bord der SOHO-Raumsonade}, Ph.D. Thesis,
  University of Kiel, 1997.

\bibitem{mueller3}
R.~Mueller-Mellin, H.~Sierks, H.~Kunow, E.~Pehlke, V.~Fleissner, E.~Rode, and
  C.~Scharmberg, ``{C}alibration of the {E}lectron {P}roton {H}elium
  {I}nstrument ({EPHIN}) {A}board the {AXAF-I S}pacecraft,''

\bibitem{vette}
J.~I. Vette, ``{AE/AP T}rapped {P}article {F}lux {M}aps 1966-1980,'' in {\em
  NSSDC Space Models Web Page},
  http://nssdc.gsfc.nasa.gov/space/model/magnetos/aeap.html, ed., 1996.

\bibitem{teague}
M.~Teague and J.~Vette, ``{T}he {I}nner {Z}one {E}lectron {M}odel {AE}-5,''
  {\em National Space Science Data Center} {\bf NSSDC 72-10}, 1972.

\bibitem{shanil}
S.~N. Virani, P.~P. Plucinsky, and R.~Mueller-Mellin, ``{U}sing {EPHIN}'s
  {D}etector {L}eakage {C}urrent {T}o {U}nderstand {R}adiation {B}elt
  {T}ransits,'' in {\em CXC Internal Memo},
  http://asc.harvard.edu/acis/radbelt, ed., 1999.

\bibitem{msm}
J.~Freeman, R.~Wolf, R.~Spiro, B.~Hausman, B.~Bales, D.~Brown, K.~Costello,
  R.~Hilmer, R.~Lambour, A.~Nagai, and J.~Bishop, ``{T}he {M}agnetospheric
  {S}pecification and {F}orecast {M}odel,'' {\em preprint} , 1995.

\bibitem{sodell2}
S.~L. O'Dell, M.~Bautz, W.~Blackwell, , D.~Brautigam, Y.~Butt, R.~Cameron,
  B.~Dichter, R.~Elsner, S.~Gussenhoven, J.~Kolodziejczak, J.~Minow, D.~Swartz,
  A.~Tennant, S.~N. Virani, and K.~Warren, ``in preparation,'' {\em Proc. SPIE}
  , 2000.

\bibitem{blackwell}
W.~Blackwell, J.~Minow, and S.~N. Virani, ``in preparation,'' {\em Proc. SPIE}
  , 2000.

\end{thebibliography}
  \bibliographystyle{spiebib}   %>>>> makes bibtex use spiebib.bst
 
  \end{document}